# INDECT Advanced Security Requirements

Manuel Urueña
Universidad Carlos III de Madrid (UC3M)
Email: muruenya@it.uc3m.es

Petr Machník
VSB - Technical University of Ostrava
Email: petr.machnik@vsb.cz

María J. Martínez
Moviquity
Email: mjmg@moviquity.com

Marcin Niemiec
AGH - University of Science and Technology
Email: niemiec@kt.agh.edu.pl

Nikolai Stoianov
Technical University of Sofia (TUS)
Email: nkl_stnv@tu-sofia.bg

***Abstract*— This paper reviews the requirements for the security mechanisms that are currently being developed in the framework of the European research project INDECT. An overview of features for integrated technologies such as Virtual Private Networks (VPNs), Cryptographic Algorithms, Quantum Cryptography, Federated ID Management and Secure Mobile Ad-hoc networking are described together with their expected use in INDECT.**

## I. Introduction

The INDECT project [1] "*Intelligent information system supporting observation, searching and detection for security of citizens in urban environment*" is a Collaborative Research Project funded by the EU 7th Framework Programme. Its main aim is to develop cost-effective tools for helping European Police Services to enforce the law and guarantee the protection of European citizens. These tools must comply with both country-level laws as well as European-level directives including, among many others, the European Declaration on Human Rights [2].

Obviously, Police information systems have stringent security requirements. In fact during the requisites gathering phase, security was deemed as the most important feature by end users, ahead of other common requisites such as performance or ease of use. Therefore, in order to guarantee the security, confidentiality and privacy of all the subsystems, users and data involved, the INDECT project has a specific task whose main objective is to supervise the security and privacy of the project [3]. Moreover, this task also carries out research on novel security mechanisms to further enhance the security of INDECT systems and other Police Information Systems. This paper is focused on the latter objective.

Although we consider that employing standard and well-studied security technologies is one of the best practices to address security, it is possible to enhance common security tools like VPNs or Key Management Centers with new components, such as a) Cryptographic Algorithms, b) Quantum Cryptography, c) Federated ID management, d) Secure ad-hoc routing protocols.

The objective of this paper is to introduce the key security components to be used in INDECT and to present the requirements of these components in a comprehensive manner.

## II. Virtual Private Networks (VPN)

Given the distributed nature of INDECT, with agents interconnected over public networks (Internet, wireless networks, etc), the use of VPNs will be required in many cases.

A Virtual Private Network (VPN) enables secure and private data transmission over an unsecured and shared network infrastructure. VPNs secure the transmitted data by encapsulating the data and then encrypting the data. Encapsulating is often referred to as tunneling because data are transmitted from one network to another, transparently across a public network infrastructure.

A good VPN solution should address most of the following issues:

- Protecting data from eavesdropping by using encryption technologies.
- Protecting packets from tampering by using packet integrity hashing functions.
- Protecting against man-in-the-middle attacks by using authentication mechanisms.
- Protecting against replay attacks by using sequence numbers when transmitting protected data.
- Defining the data encapsulation and protection mechanisms, and how protected traffic is transmitted between devices.
- Defining what traffic is necessary to protect.

Not every VPN implementation should include all of these components or does not implement them as securely as other VPNs. Therefore, it is very important to define a security policy in order to determine what VPN technology is the most suitable one for a particular situation. It is often possible to use multiple VPN solutions in the same network. Nowadays, the most important types of VPN technologies are IPSec, SSL VPN, GRE, MPLS VPN, PPTP and L2TP.

### A. VPN requirements

To choose the most suitable VPN solution, it is necessary to consider the following criteria [4]: Security, Implementation, Management, Support, High Availability, Scalability, Flexibility and Cost.

*1) Security:* First of all, it is necessary to determine what data needs to be protected. If it is necessary to protect traffic for specific applications, such as e-mail, web, file transfer, and others, the best solution can be a SSL VPN. If it is necessary to protect all traffic for specific hosts or network segments, it is probably better to use another type of VPN like IPSec. Second, it is necessary to consider what kind of protection is required: encryption, packet integrity, authentication. And third, how much protection is needed? This means that it is necessary to choose a sufficiently strong encryption algorithm (e.g. 3DES, AES) and a reliable authentication method (pre-shared keys, digital certificates) for the desired security level.

*2) Implementation, Management and Support:* Different VPN solutions have different demands on implementation, management, and support. For example, SSL VPNs are easier to set up, manage, and troubleshoot compared to IPSec. Regarding key management, certificates from a PKI are usually easier to manage and distribute than pre-shared keys.

*3) High Availability:* Some VPN implementations support redundancy and some others don't. It is also necessary to determine the type of redundancy that is needed - chassis or connection redundancy - and how well different VPN implementations can deal with the specific redundancy method. Moreover, some networking vendors implement other proprietary redundancy features.

*4) Scalability and Flexibility:* When choosing a VPN implementation, it is convenient to ensure that the chosen solution is scalable and flexible. The solution needs to be scalable to accommodate the future growth of the network, and flexible to deal with changes that occur within the network. In other words, if it is needed to add more sites to the VPN design, it is necessary to consider how much work it is required to perform this change, how many devices are needed to configure or reconfigure, how much configuration has to be performed on these devices, whether additional overhead can affect existing devices, etc.

*5) Cost:* Some costs that need to be evaluated are:

- Hardware devices.
- Software products and licenses.
- Network vendor maintenance and support costs.
- Personnel costs and personnel training costs.
- Bandwidth for the VPN traffic and associated overhead.

### *B. Future directions*

Because INDECT systems will exchange sensitive data in many ways, it will be necessary to choose the most convenient VPN solutions for each individual case. The most suitable solution for the remote access of Users to a private network and for the web-based data transmission seems to be the SSL VPN technology. On the other hand, to secure the entire communication among remote private networks over an insecure network, it would be better to use the IPSec VPN technology. It will be also desirable to consider the different security levels of chosen VPN solutions regarding used cryptographic algorithms, authentication and key distribution methods.

## III. Cryptographic Algorithms

One of the main research areas addressed by INDECT is cryptographic algorithms development. This section reviews security basics of cryptographic algorithms, symmetric and asymmetric ciphers as well as key management issues, in order to properly define the requirements for the new cryptographic algorithms to be developed.

### *A. The basics of cryptographic algorithms*

Cryptographic algorithms are mainly applicable for providing data confidentiality and integrity, as well as identification and authentication processes [5]. The cryptographic algorithm (cipher) is a mathematical function which transforms data to an unreadable form to anyone except the key owners. Keys are binary strings which are used for plain-text encryption and ciphered text decryption. In general, we can specify two types of cryptographic algorithms: symmetric and asymmetric.

### *B. Symmetric ciphers*

In symmetric algorithms the encryption key can be calculated from the decryption key and vice versa. In most cases the encryption and decryption keys are exactly the same. Symmetric algorithms, which are also called secret key algorithms, require that both the sender and the recipient have exchanged in advance their respective keys, and they are responsible to keep them in secret. The security of the symmetric algorithms is based on the key. If the secret key is found, everyone is able to encrypt and decrypt any message.

*1) DES and Triple DES (3DES):* The Data Encryption Standard (DES) was a well known block cipher operating on 64-bit data blocks. The encryption transformation depends on a 56-bit secret key and consists of sixteen Feistel iterations surrounded by two permutations.

Since a 56-bit secret key is considered insecure, Triple DES was later proposed. The idea behind Triple-DES is to increase the number of DES encryptions three times with two or three different keys. This method gains strength both against cryptanalytic attacks as well as against exhaustive search. However, it is weak against related key attack, and the speed is three times slower than single-DES.

*2) Advanced Encryption Standard (AES):* AES, originally published as Rijndael, is a block cipher with a variable block size and variable key size. The key size and the block size can be independently specified to 128, 192 or 256 bits. In the simplest scenario we have 128-bit key size and the same block size. In this case, a 128-bit message (plain text, cipher text) block is segmented into 16 bytes. AES algorithm was designed to have the following characteristics: resistance against all known attacks, speed and code compactness on a wide range of platforms, as well as design simplicity.

### *C. Asymmetric ciphers*

In asymmetric algorithms, which are also called public key algorithms, the encryption and decryption are performed with two different keys: a private key and a public key. In addition to this, the decrypting key cannot be calculated from the encryption key. The encryption key is called public and it is not necessary to be kept in secret. Everyone can use it for encryption, but only the one who has the respective decryption key can decrypt the message. The decryption key is thus called personal or private. By means of asymmetric ciphers we can encrypt messages, as well as to authenticate a user in possession of the private key. The most popular public-key cryptosystem is RSA, named after its inventors Rivest, Shamir and Adleman.

### *D. Key management*

The security of one symmetric cryptographic system is a function of strength of the algorithm and length of the key. In real life the key management of a cryptographic system is one of the most important aspects of its security [6]. The introduction of a secure management of keys is crucial, and quite often the attacks to cryptographic systems are directed to the weaknesses of key management. In computer systems the key management is normally done by means of a Key Management Center. Key management implies: generation, distribution, use, storage, time (period) of use, change and destruction of keys.

Taking into account the complexity and the broad scope of the INDECT project, as well as the big number and variety of end-users devices, we consider that the most favourable approach for creation of cryptographic Key Distribution and Management Center is a combined method, thus the center should support both symmetric and asymmetric algorithms. Additional functionality could be digital certification. It is worth to mention that some devices, i.e. video cameras, support this technique.

A brute-force attack (check of all possible keys) is the last available technique for breaking the cryptographic system. Two parameters define the time of a successful brute-force attack: number of keys to check and speed (time) of each check. The selection of the key length for the specific application depends on the time we want the data to be protected and on the value of those data.

### *E. Future directions*

INDECT will develop some new cryptographic algorithms based on S-Boxes to ensure the security of INDECT system. Below we present crucial requirements and features of the new solutions.

- The cipher algorithms employed to ensure data confidentiality must be strong and must not have any backdoors allowing decrypting data without the key. The minimal key size should be 128 bits for symmetric algorithm and 2048 bits for asymmetric algorithm.
- Cryptographic algorithms should support a shared key service to ensure multi-user encryption.
- The security of a cipher must be based on the strength of the algorithm (must not be based on the secrecy of the algorithm).
- Keys should be generated by means of random number generators. These keys should be statistically tested to ensure truly randomness.

## IV. Quantum Cryptography

Quantum Cryptography (QC) is other topic of INDECT research on security. QC is a powerful security technique which provides unbreakable communication via optical or wireless links. Usually, it is based on phase-encoding of the photons that are transmitted over a dedicated optical fibers, called "dark fibers". QC ensures the highest security level, because it is not possible to eavesdrop the communication in a passive way. If an eavesdropper reads transmitted data, it will change the quantum states of the photons and he will be discovered [7].

The Quantum Key Distribution (QKD) protocols could be based on phase-encoding of the photons (single particles) or based on quantum entangled pairs of photons. An entangled photon pair is a couple of photons with a unique correlation, such that the property of one photon is automatically and instantly determined at the moment the other photon's property is measured. Nowadays, a lot of QKD algorithms (i.e. BB84, E91, B92,SARG04) have been proposed in the literature. Moreover many different QC devices (e.g. produced by idQuantique, SmartQuantum, MagiQ) have been already constructed. Because QC is moving outside the research laboratories, it is now possible to apply this technique in practice. The crucial issue is the integration of QC with existing networks, because building additional optical infrastructure in metropolitan areas is very expensive.

It is still not clear if QKD will cause a revolution in communications. Although there are still a lot of problems with deploying the QKD in real optical or wireless communication networks, we can exchange information secured by means of QC in specific environments. One of the crucial issues concerns requirements. QC in optical networks will be possible if we solve a few practical problems and meet some specific requirements. They could be divided into two groups: end-user requirements and system requirements [8].

### *A. End-user requirements*

At the beginning of the system design process, end-users should define their specific requirements. This requirements should come from all stakeholders and express the desired properties of the whole communication system. When we consider the QKD technique, the early adopters of systems that will be protected by means of QC are: banks, big corporations or public administration.

*1) Security:* Very often data security is a crucial issue. Let's consider a communication system of the Police. A weak security system jeopardizes the Police to carry out cover investigations, but is could even put in risk the life of many people. QC ensures the highest level of system security.

*2) Cost:* The cost is usually the most frequently end-user requirement that appears during the design process. Unfortunately, QC is a very expensive technique. Therefore, only when end-user requirements do not include severe cost restrictions, we can seriously think about deploying QC.

*3) Performance:* Performance is a feature of the system that is inseparable from security, because it is usually inversely proportional to the security level of the system. QC in comparison with asymmetric ciphers provides much more efficient data encryption, because enables secure secret key exchange prior to data transmission.

*4) Ease of use:* End-users of communication systems usually are not engineers and do not possess advanced technical skills. Fortunately QKD devices deployed in communication system do not generate additional difficulty in user interface. The objective of this technique is to be completely 'transparent' for end-users.

### *B. System requirements*

The second group of requirements - System Requirements - express desirable properties of the system itself. They are important because, on the one hand, system requirements should lead to the achievement of user requirements and, on the other hand, should indicate how to archive the optimal solution from the available technology point of view.

*1) Confidentiality:* The best way to ensure confidentiality is data encryption. Currently, network communications are often protected using a symmetric encryption algorithm, like AES or 3DES. When data confidentiality is the crucial requirement, QC in conjunction with strong symmetric cipher (e.g. AES with a key size of 256 bits) is really a good solution.

*2) Communication security:* Communication security ensures that information flows only between the proper end points. QC ensures this requirement by means of the principles of quantum mechanics.

*3) Privacy:* Privacy assurance is the protection of user data against disclosure. When eavesdroppers want to observe the quantum information, they change the quantum states and can be discovered.

*4) Scalability:* This feature of the system does not require to change the whole security system but only an extension of the system when a bank or other big corporation has to connect a new branch. QC is thus a scalable technique.

### *C. Future directions*

Currently we are developing a high-level QC security protocol to ensure the highest level of data security. Below we present security requirements for quantum cryptography. We should meet these requirements to be sure that the implemented quantum cryptography protocol is secure enough for the target purpose.

- Quantum information should be coded by means of single photons to avoid split light and receive information in secret way, and should be transmitted using dedicated fibers ('dark fibers').
- High-level QC protocols must compare enough bits to be sure that nobody has eavesdropped the communication.
- Cipher algorithms that are used in quantum cryptography must ensure high level of confidentiality and must use long and randomly generated keys to encrypt data.

## V. Federated ID Management

A Federated Identity Management system is being developed in the INDECT project. It is tailored to support the coordination of different departments and security forces, with the provision of a simple authentication mechanism that controls the access to the corresponding inter-connected systems. Along with content heterogeneity, due to the multiple data sources of sensitive information from the different project subsystems, comes the need to support multiple access levels depending on the user roles/attributes. The federated management system will operate across existing technical boundaries including heterogeneous operating systems and security platforms [14].

### *A. Identity Management in INDECT project*

Layered web authentication and access control services should be implemented in the two main interfaces with the end-users (e.g Police Officers). Policemen will use their PCs and interact with the integrated INDECT Portal website while working in the office, and during field duty they will use their mobile equipment, that is, PDAs/Smartphones. Federated ID will provide them with a highly ergonomic (contextual, collaborative and personalized) single point of access to multiple data sources. However, given the heterogeneity of the information being handled, there will also be an Advanced Interactive Video Streaming System (IVAS) as an alternative channel to support communication and processing of multimedia streams. The underlying interoperable web-based SOA architecture will be prepared for two different profile categories in the access management: either for passive clients (web browsers), using an HTTPS connection, or active clients (smart/rich apps), with SOAP/XML messaging. Access decisions will be based on the roles that individual users have as part of the Police organization.

### *B. Authentication and Authorisation methods*

Within a single organisation, Single Sign On (SSO) could be a suitable way to authenticate the users in a variety of disparate systems in order to give them access to multiple resources using a single set of credentials. The user would log into a client or terminal using the corresponding SSO-based username and password, and then the system would validate its authenticity and log the user into the underlying systems in a transparent way (username and password stored in a secure directory).

However, for a multi-organisation network, a complete Federated ID management system should be the security solution to guarantee that the system being accessed is the intended one and that the user attempting to log in is who she claims to be. The procedure is based on both need-to-know and the role of the individual or group in the process requiring access. In this case, the access would no longer be based on relatively static organisational structures but would be built from dynamic, process-based operational/functional requirements, which in this case are tailored to the INDECT context of use, enabling the horizontal fusion of the distributed services provided through the different INDECT subsystems.

### C. Federated ID Specifications

Federated Identity management is the process required to establish secure and cost-effective business collaboration with an auditable trail of who is connecting to the target application and resources [13]. It would allow individuals to use the same personal identification to have access to a 'circle of trust/confidence', thus being able to share information between domains that have their own directories, security or authentication technologies. Efficiency in submission of access control details lowers the risk of security threats and identity management costs, while also ensuring the interoperability between business partners.

Federated infrastructures allow having a unique identity, hosted in an Identity Provider (IdP) or account partner, that is accepted by one or more Service Providers (SP) or relying parties. The latter allows a user to access a resource or an application; and the former is responsible for user authentication. The system formed by one or more IdPs and SPs is called a 'federation', and it is characterized by having a relationship of trust among its members, simplifying data communication and validation of users in a secure way. Trust management addresses relationships (system-to-system, business-to-business) between entities within organizations, security domains and systems, dealing with business and technological aspects altogether.

Identity details and authentication complexity could be hidden behind the protocols and APIs elaborated in different specifications. Some well known Federated Identity standards that have been developed for enterprise-centric models are Security Assertions Markup Language (SAML), Liberty ID-FF and Web services (WS) Federation:

- OASIS SAML TC [12]: Specs based in XML that allow cross-domain authentication. Two main components define SAML assertions, describing security tokens representing users, and SAML bindings and profiles. SAML protocol has three versions: SAML 1.0, 1.1, (1.x for single sign-on functionality) and 2.0, which takes input from Liberty Identity Federation Framework (ID-FF) 1.2 but with a major functional increase (a deeper consideration of the identity life cycle functionality and privacy concerns).
- Liberty Alliance Project [9]–[11]: Federated single sign-on protocols developed by a group of vendors (i.e. IBM) as part of the Liberty Identity Federation Framework (ID-FF). There have been Liberty ID-FF 1.1 and 1.2 releases, the latter being superseded by SAML 2.0.
- WS-Federation [15]: Extensions to the WS-Trust Security Token Service to enable standard based security approaches for both browser-based and Web services-based applications in service-oriented architecture (SOA) environments. It provides to federation partners: identity brokering, attribute request and retrieval, and authentication and authorization claims with privacy protection. There have been two releases of WS-Federation: 1.0 and 1.1.

In spite of the limitations associated with Federated ID, mainly referred to its complexity and constrains for deployment (due to the lack of support by governmental organizations and the non-existence of an underlying legislation), this access control solution is proposed as the most suitable one in order to centralize the authentication of INDECT subsystems.

### D. Future directions

Technical solutions for Federated ID in European ID authentication and Identity Management require the analysis of the limitations of current federated identity approaches and technologies in order to cover the European citizens requirements and expectations for non-anonymous services. Most of them refer to the incorporation of more partners and to the exploitation of more technical choices for monitoring and surveillance in virtual or real environments targeted to protect the citizens. Intended users at later stages of the INDECT project will be all the services and institutions responsible for crime protection and the maintenance of public security (departments of public prosecutions and justice and also counter-terrorism). Furthermore, new standards-based Web services with increased capability (improved interface, greater flexibility and reliability) may be made available to this wider circle of trust, considering their requirements for security and privacy. In this framework for added-value services and seamless access to a number of departments, any kind of identity management duplication should be avoided. Sharing identities across the borders will require, at least, the use of common standard specifications and unified definition of profiles (i.e. categories for user rights or procedures for account linking and alias management). New technologies, knowledge and methodologies will also need to be developed in an interoperable framework (i.e. more sophisticated human recognition systems, separately watermarked components, etc).

To sum up, the future projection of Federated ID will require the management of an increasing number of functionalities provided by means of web services. Process reuse adapted to the business logic in autonomous distributed services will be the basis in the standards. The awareness of the technical security vulnerabilities due to seamless integration will lead the project to consider an Identity Management framework supporting the effective control of user identities, establishing trust, enabling privacy, and complying with legal and regulatory environments.

## VI. Mobile Security

Wireless Ad-hoc networks are data networks that do not rely on fixed infrastructure, but they are built automatically among the devices in a certain area. Therefore they are a very appealing tool for security forces, especially to manage emergencies when traditional communication networks do not work or are collapsed. However ad-hoc networking poses a number of challenges in order to provide truly auto-configuration, user mobility, robustness, performance, energy efficiency and confidentiality. In particular, confidentiality is a must for an ad-hoc network for security forces. Although there are plenty of security mechanisms for computer networks, security in ad-hoc networks is still an open issue because of the lack of access to a trustable third party.

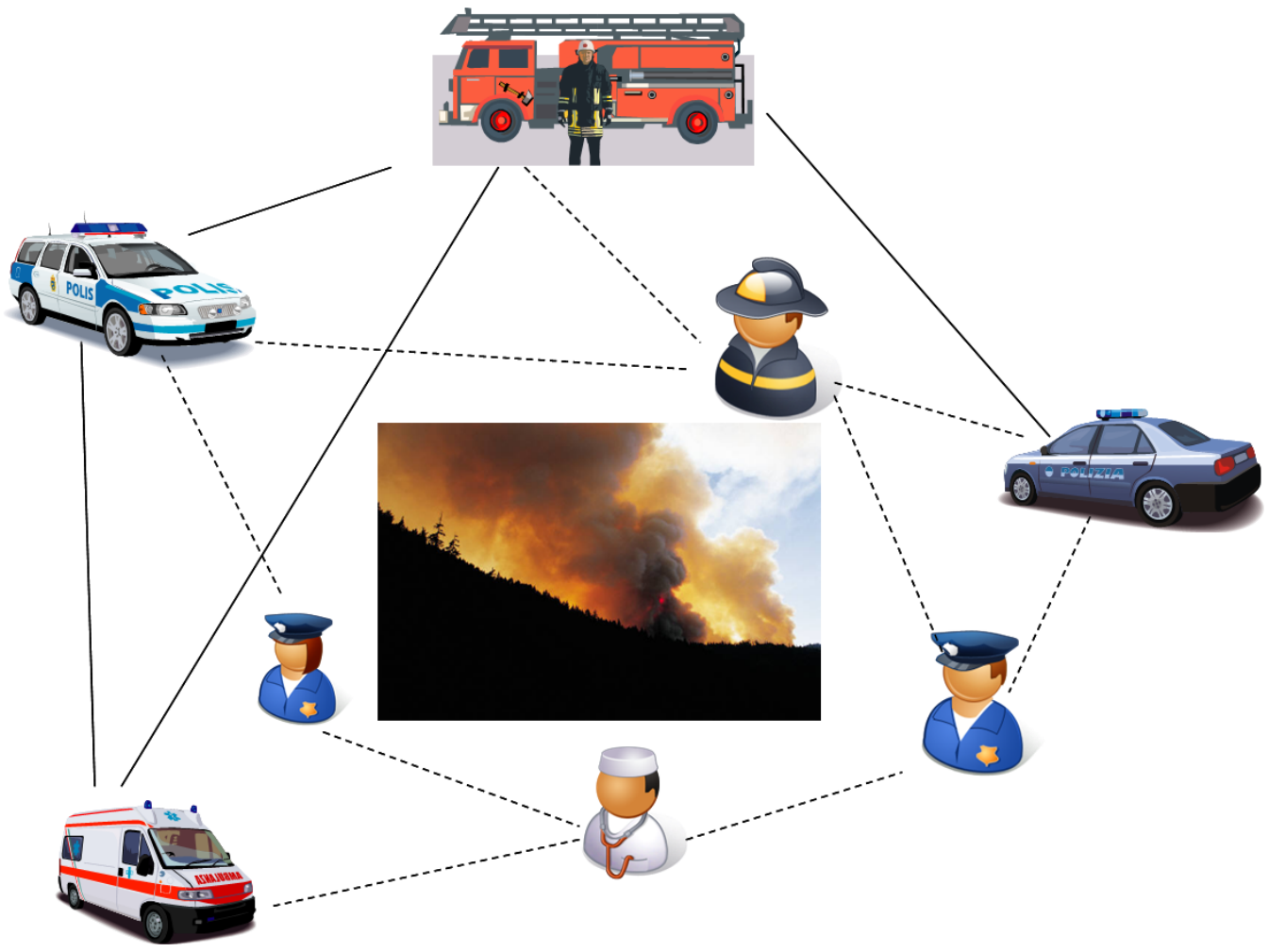

Fig. 1. Multi-organization Ad-hoc network for emergency response.

### A. Ad-hoc Security

In order to enhance the security of an ad-hoc network, several layers of the TCP/IP stack should be considered. Application layer should deal with end-to-end security issues, whereas ad-hoc routing and lower layers lack of most of the security mechanisms developed for standard TCP/IP stacks.

In particular, the application layer (or a session layer tightly coupled with it) should provide security at Users level. For instance provide Authentication of the other communicating peer, Confidentiality, Integrity, and non-Repudiation depending on the application. From a communication stand-point most of these problems can be solved with well know techniques such as TLS/SSL and User certificates. However, it is necessary to consider the especial characteristics of this scenario, namely the lack of fixed infrastructure and the need to locate nearby users from different, not fully trusted organizations.

Probably, designing a secure ad-hoc routing is the most challenging aspect, since ad-hoc routing is still in its infancy, moving from the research labs to initial large-scale deployments. For instance, there is not a single standard ad-hoc routing protocol but many, tailored to specific scenarios. Reactive protocols, like AODV [18] or DSR [19], are best suited for mobile nodes, whereas Pro-active ones, like OLSR [17], have been mostly deployed in mesh networks with little mobility. Probably the most useful scenario for security and emergency forces is the so called Mobile Ad hoc Network or MANET, where most nodes are mobile and relationships among nodes do not last long. For this reason we will focus on the security of reactive protocols, especially AODV and DSR, since they are the most popular routing protocols for MANETs.

Although there are some works in the literature about security in ad-hoc networks, there is still little experience with real wireless ad-hoc networks and far less applied security research when compared to their fixed counterparts. Most works [16] are more focused on identifying a particular attack for a given protocol and its specific countermeasure, than providing a global security solution. This is a brief list of the identified attacks (some of them are generic, whereas others only affect a particular routing protocol):

- *Black/Gray hole attack*. Attract all traffic towards an attacker node by means of routing, and then drop it all or just a fraction of it in order to avoid detection.
- *Sleep deprivation attack*. Redirect traffic continuously towards a victim in order to deplete its battery.
- *Rushing attack*. Destination nodes usually reply only to the first RREQ message. Thus attackers send their own RREQ message before others to impose a given route.
- *Modify Hop/Sequence number* (AODV only). Attacker could promote himself by advertising a route with less hops or a greater sequence number.
- *Modify Source Route* (DSR only). DSR specifies the full path in the RREP messages. Thus an attacker could modify the RREP route at its will.
- *Broadcast false routes* (DSR only). In order to save bandwidth DSR nodes have a cache that is populated by overhearing RREP messages. An attacker could poison DSR route caches by generating false RREP messages.
- *Falsify Route Errors*. An attacker could generate a DoS attack by tearing down valid routes.
- *Selfish/Malicious MAC layer*. An attacker may configure its wireless interface with lower back-off values in order to gain some performance advantage.
- *Jamming*. An attacker may generate a strong signal that interferes with the wireless channel being employed.

There are a number of solutions for some of the above attacks, although most works usually focus just in one of them:

- *Encrypt and Sign all messages*. To avoid eavesdropping and message modification in transit. However this solution requires some trusted key distribution mechanism.
- *Access Control*. Exclude attackers from routing by authenticating new nodes before joining to the MANET.
- *Reputation mechanisms*. When no authorization is possible, some authors propose to build a reputation mechanism in order to classify nodes into cooperative or uncooperative classes.
- *Levels of security/trust*. In order to limit insider attacks, the routing protocol prefers nodes from the same security level (e.g. Police officers) than just choosing the shortest path.
- *Randomize Message Forwarding*. To avoid rushing attack, some authors propose to queue and send RREQ messages at random times. However this may add extra latency to path discovery.
- *Do not use hop count*. Since hop count field can be forged

by intermediate nodes, ignore it and just reply to the first RREQ.
- *Hash chains*. An alternative to secure the hop count is to generate a hash chain so a shorter path cannot be claimed.
- *Onion routing*. Encrypt the DSR source route between hops to avoid tampering.
- *Overhearing / Probing / Feedback*. To detect packet dropping by means of a black/grey hole attack.

### B. Future directions

After analyzing the literature about security on ad-hoc networks, we consider that low-level attacks, such as jamming and non-cooperative CSMA/CD medium access, are the most important threats in ad-hoc networks because it is quite difficult, if not impossible, to defend the network from these kinds of attacks (i.e. requires specialised hardware). Therefore it is better to assume that an adversary is able to disrupt any wireless link within a certain range, and design a routing protocol that is able to detect this damaged zone and quickly route around it. In particular we are investigating how a routing protocol with multi-path capabilities, which is able to discover/employ different paths in parallel, can be employed in this case to limit the impact of a lower-layer attack and to quickly recover from it.

## VII. Conclusion

Nowadays security is one of the most important features of a communications system. This is specially true for Police information systems such as the ones being developed by the INDECT project. This paper has reviewed a few standard and under-study security mechanisms to be deployed to support the security of INDECT systems. In particular a full categorization of Virtual Private Network (VPN) technologies has been presented, as well as key aspects of the research on novel Cryptographic Algorithms based on S-Boxes, and a new Quantum Cryptography (QC) protocol for top-security systems. The creation of a Key Management Center and the usage of Federated ID technology are key aspects in order to glue together all the heterogeneous information sources and INDECT systems in a secure way. Finally, a review of the state of the art in Mobile Ad-hoc Networks and the rationale behind a new multi-path ad-hoc routing protocol to withstand low level attacks (e.g. jamming) are also presented.

## Acknowledgment

The authors would like to thank David Larrabeiti for his comments. The work presented in this paper has been funded by the INDECT project (Ref 218086) of the 7th EU Framework Programme.